\documentclass[useAMS,usenatbib]{mn2e}

\usepackage{graphicx,natbib}
\usepackage{amsmath}
\usepackage{subfigure} 
\usepackage{hyperref}
\usepackage{amssymb}

\title[Vertical  waves  induced by  minor  mergers] {Vertical  density
  waves in the Milky Way disc induced by the Sagittarius Dwarf Galaxy}

\author[F.~A. G\'omez et al.]{Facundo A. G\'omez$^{1,2}$\thanks{Email:fgomez@pa.msu.edu},  
Ivan Minchev$^{3}$, 
\vspace{0.05cm}
Brian W. O'Shea$^{1,2,4,5}$,
Timothy C. Beers$^{6,1,5}$,\newauthor 
James S. Bullock$^{7}$,
Chris W. Purcell$^{8}$
\vspace{0.2cm}
\\
$^{1}$ Department of Physics and Astronomy, Michigan State University, East Lansing, MI 48824, USA\\
$^{2}$ Institute for Cyber-Enabled Research, Michigan State University, East Lansing, MI 48824, USA\\
$^{3}$ Leibniz-Institut f\"{ur} Astrophysik Potsdam (AIP), An der Sternwarte 16, D-14482, Potsdam, Germany\\
$^{4}$ Lyman Briggs College, Michigan State University, East Lansing, MI 48825, USA\\
$^{5}$ Joint Institute for Nuclear Astrophysics (JINA), Michigan State
University, East Lansing, MI 48824, USA \\
$^{6}$ National Optical Astronomy Observatory, Tucson, AZ 85719, USA\\
$^{7}$ Department of Physics \& Astronomy 4129 Frederick Reines Hall, University of California, Irvine, CA 92697, USA\\
$^{8}$ Department of Physics and Astronomy, University of Pittsburgh, Pittsburgh, PA 15260, USA\\
}

\begin{document}

\date{}

\pagerange{\pageref{firstpage}--\pageref{lastpage}} \pubyear{}

\maketitle

\label{firstpage}

\begin{abstract}

  Recently,  Widrow  and  collaborators  announced  the  discovery  of
  vertical density waves in the Milky Way disk.  Here we investigate a
  scenario  where these waves  were induced  by the  Sagittarius dwarf
  galaxy  as   it  plunged   through  the  Galaxy.    Using  numerical
  simulations,  we  find  that  the the  Sagittarius  impact  produces
  North-South  asymmetries   and  vertical  wave-like   behavior  that
  qualitatively  agrees with what  is observed.   The extent  to which
  vertical  modes can  radially penetrate  into the  disc, as  well as
  their amplitudes,  depend on the  mass of the  perturbing satellite.
  We show  that the mean height of  the disc is expected  to vary more
  rapidly  in the  radial than  in  the azimuthal  direction.  If  the
  observed  vertical density  asymmetry is  indeed caused  by vertical
  oscillations, we predict radial and azimuthal variations of the mean
  vertical  velocity, correlating  with the  spatial  structure. These
  variations can have amplitudes as large as 8 km s$^{-1}$.

\end{abstract}

\begin{keywords}
  Galaxy:  disc,   structure  --  galaxies:   formation  --  galaxies:
  kinematics and dynamics -- methods: $N$-body simulations
\end{keywords}

\section{Introduction}

Minor mergers can significantly perturb the overall structure of their
host  galactic  disc  \citep[][]{quinn93,alvaro08}.   As  they  merge,
relatively small satellite galaxies can induce the formation of spiral
arms   and  ring-like   structures  as   well  as   radial  migration,
significantly flare  or warp the disc,  and influence the  growth of a
central                                                             bar
\citep[][]{tutu,kaza08,alvaro08,younger,m09,quillen09,p11,bird,g12b}.

\citet[][hereafter P11]{p11} presented  simulations of the response of
the  Milky Way  (MW) disc  to tidal  interaction with  the Sagittarius
dwarf galaxy (Sgr).  They showed that many of the global morphological
features observed in the Galactic disc can be simultaneously explained
by this  interaction.  An example is the  kinematically cold structure
known as the Monoceros ring \citep{newberg02,juric08}, which naturally
emerges in these simulations, although its origin is still a matter of
debate \citep[see][]{conn,linew,mono}.  \citet[][hereafter G12]{g12a} showed that
perturbations  observed in  the phase-space  distribution of  old disc
stars in the Solar  Neighbourhood (SN) can be reproduced qualitatively
with  these  simulations.   They  interpreted  such  perturbations  as
signatures of \emph{radial} density waves  excited on the plane of the
disc by Sgr.  Perturbations  in the \emph{vertical} direction were not
explored in their  work. However, using a much  larger photometric and
spectroscopic   data   set,   \citet[][hereafter  W12]{w12}   recently
identified a North-South asymmetry in both the spatial density and the
velocity distribution  of SN stars.  The asymmetry  has the appearance
of  a coherent, wave-like  perturbation, intrinsic  to the  disc.  W12
speculate  that  this perturbation  could  have  been  excited by  the
passage  of a  satellite galaxy  through  the Galactic  disc. In  this
Letter,  we  explore  the  possibility  of  Sgr  being  the  perturber
associated   with   {\it  both}   the   vertical   and  radial   modes.

\section{Simulations}
\label{sec:sim}

In  this Section  we briefly  describe our  simulations; we  refer the
reader  to  P11  and  G12   for  a  more  detailed  description.   Two
simulations  with different  models for  the Sagittarius  dwarf galaxy
progenitor were performed.  A Light (Heavy) {\it Sgr} progenitor, with
effective  virial  mass $M_{\rm  vir}  =  10^{10.5}~{\rm M}_{\odot}  ~
(10^{11} ~ {\rm  M}_{\odot})$, was initialised with a  NFW dark matter
halo  of scale  length 4.9  kpc  (6.5 kpc),  self-consistently with  a
separate stellar component.  For the stellar component, a King profile
with core radius  1.5 kpc, tidal radius 4 kpc,  and a central velocity
dispersion of 23~km s$^{-1}$ (30 km s$^{-1}$) was used. The satellites
were launched 80~kpc from the Galactic  centre in the plane of the MW,
traveling  vertically  at 80~km  s$^{-1}$  toward  the North  Galactic
Pole. The  mass loss  that would have  occurred between  virial radius
infall and  this ``initial'' location  is accounted for  by truncating
the progenitor DM halo mass  profile at the instantaneous Jacobi tidal
radius, $r_{\rm t} = 23.2$ (30.6)  kpc. This leaves a total bound mass
that is  a factor of $\sim  3$ smaller than the  effective virial mass
originally   assigned.    The    simulations   reach   a   present-day
configuration after $\approx 2.7$ Gyr (2.1 Gyr) of evolution.  In both
cases, the  host galaxy includes a  NFW dark matter halo  with a scale
radius $r_{\rm  s} =  14.4$~kpc and virial  mass ${\rm M}_{\rm  vir} =
10^{12} ~  {\rm M}_{\odot}$; the disc has  a mass of $3.59  ~ \times ~
10^{10} ~ {\rm  M}_{\odot}$, an exponential scale length  of 2.84 kpc,
and a vertical scale height of  0.43 kpc; the central bulge has a mass
of  $9.52 ~  \times ~  10^{9} ~  {\rm M}_{\odot}$  and an  $n  = 1.28$
S{\'e}rsic profile, with an effective radius of 0.56 kpc.  The initial
disc in both Sgr-infall models is completely smooth at $t=0$ Gyr.  The
simulations followed the evolution of 30 million particles with masses
in the range 1.1 - 1.9 $\times ~ 10^{4} {\rm M}_{\odot}$.

\section{Perturbations in local volumes}
\label{sec:local}

\begin{figure}
\centering
\includegraphics[width=85mm,clip]{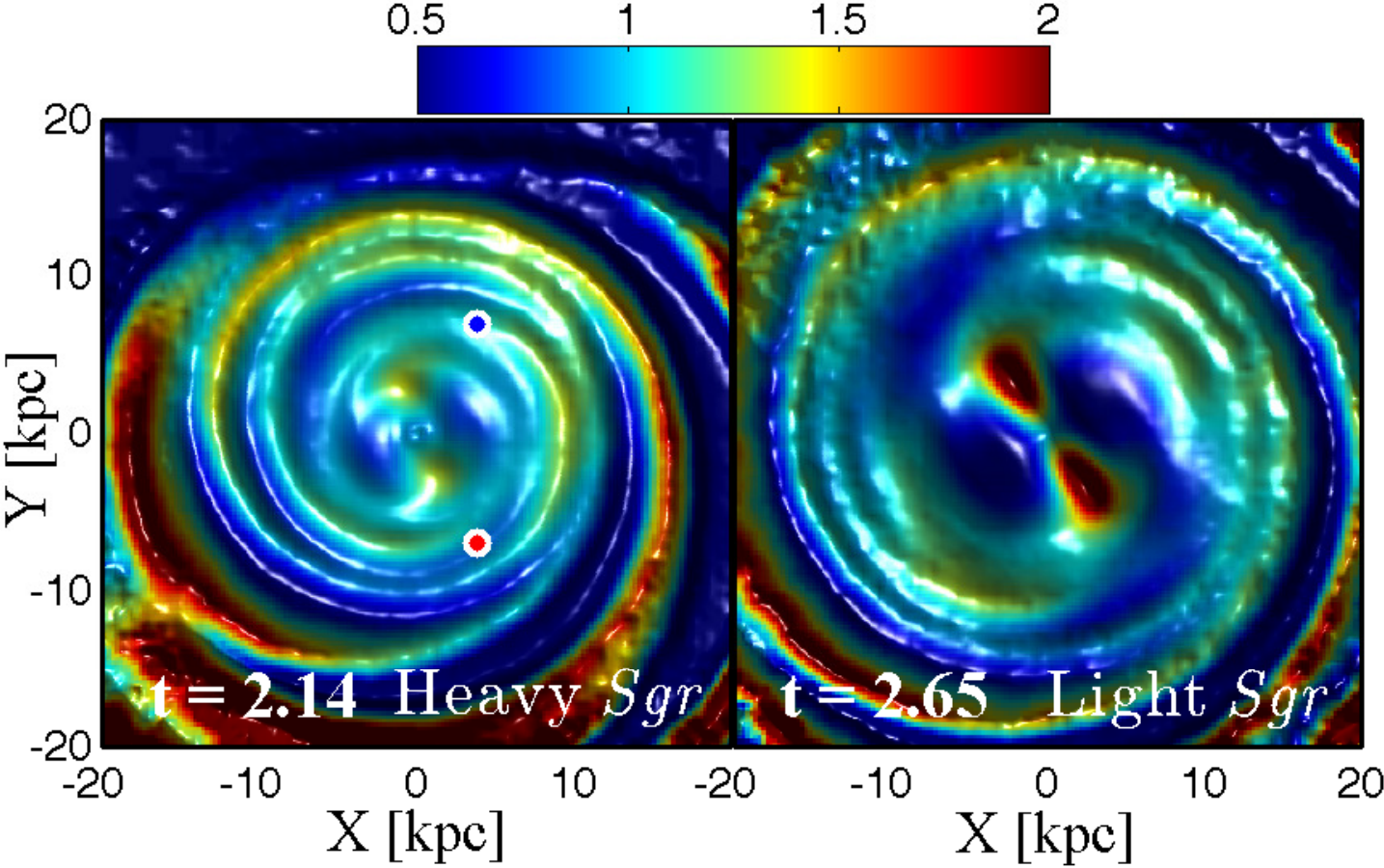}
\caption{Overdensity map  of the Heavy  (left) and Light  (right) {\it
    Sgr} simulations at the  present time. The colours and relief indicate
  the ratio of the local stellar density to the mean axysimmetric disc
  density at the corresponding  galactocentric distance.  The blue and
  red  dots indicate  the location  of  the volumes  explored.}
\label{fig:dens}
\end{figure}

\begin{figure}
\includegraphics[width=83mm,clip]{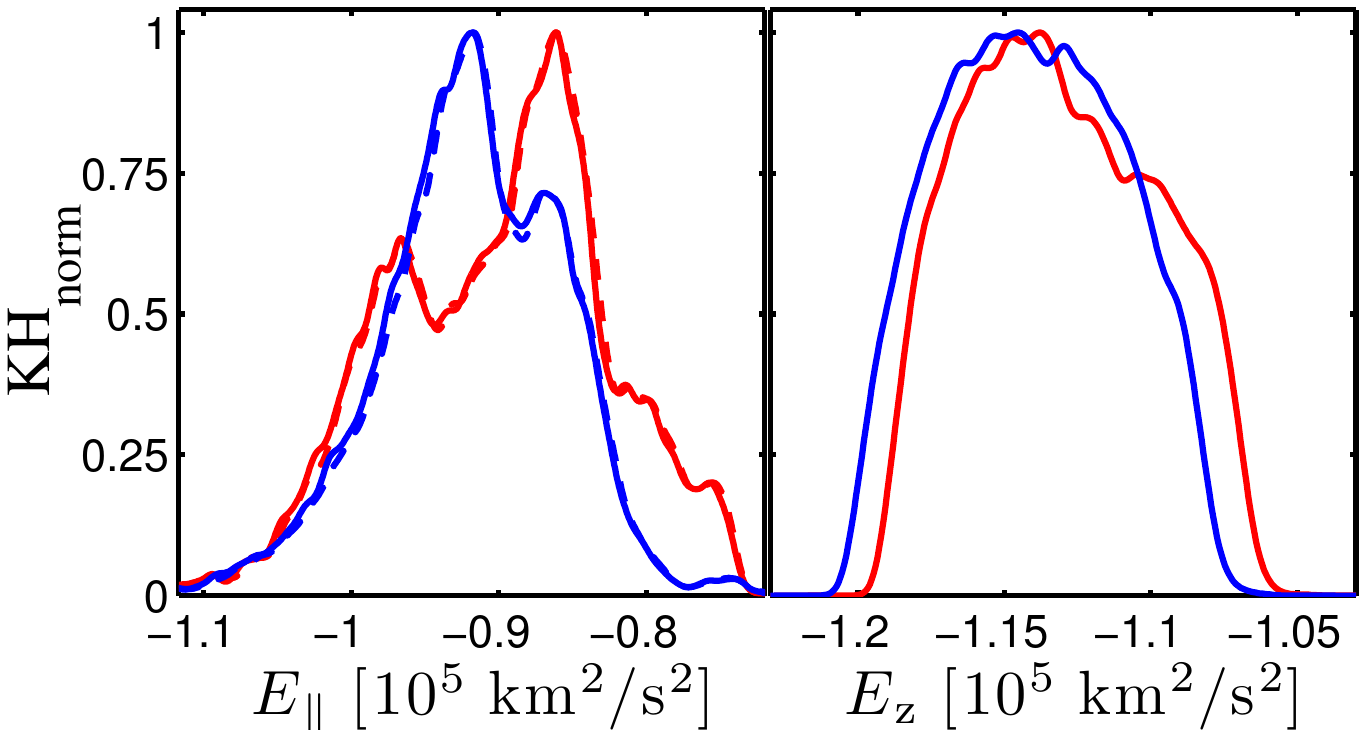}
\caption{{\it Left panel}:  Total (dashed lines) and  in-plane (solid lines)
  energy distributions  of disc particles  located within the  blue and
  red volumes shown in Fig~\ref{fig:dens}. {\it Right panel}: As in the left
  panel, but for the vertical energy distribution.}
\label{fig:ener}
\end{figure}

Fig.~\ref{fig:dens} shows  an overdensity map of the  Heavy (left) and
Light (right) {\it Sgr} simulation discs at present-day configuration.
The maps are obtained by  normalising the local stellar density to the
mean   axysimmetric  density   at  the   corresponding  galactocentric
distance.  The non-axysimmetric energy  kick imparted by the satellite
as it  merges with  the host induces  the formation of  spiral density
waves. As shown by G12, waves excited by the Heavy {\it Sgr} satellite
can  be detected  in  SN-like volumes  as  peaks in  the local  energy
distribution.  These density waves are mainly in the radial direction.
In fact, in very small local volumes (i.e., distances $\leq 0.2$ kpc),
these waves  can be  observed as well-defined  features in  the radial
($v_{r}$)     and    tangential     ($v_{\phi}$)     velocity    field
\citep[see][G12]{m09}.  In  the left panel  of Fig.~\ref{fig:ener}, we
show with dashed  lines the normalised total energy  distribution $E =
f(\mathbf{x},\mathbf{v})$ of two SN-like cylindrical volumes extracted
from the Heavy {\it Sgr} simulation.  The volumes have a 1 kpc radius,
and are  located at 8 kpc  from the galactic  centre. Their locations,
chosen based on  G12 results, are indicated with a blue  and a red dot
in the left panel of Fig.~\ref{fig:dens}.  Note the well-defined peaks
in these  distributions, which reveal  the presence of  density waves.
As  expected  from  disc  orbits,  the  in-plane  energy  distribution
$E_{\parallel} =  f(\mathbf{x},v_{r},v_{\phi})$ (solid lines)  is very
similar to the total  energy distribution, indicating that these peaks
are associated  with perturbations  mainly in the  plane of  the disc.
Nevertheless,   the  stellar  particles   in  these   volumes  present
non-symmetrical energy distributions in the vertical direction $E_{\rm
  z}  = f(\mathbf{x},v_{\rm  z})$,  as  shown in  the  right panel  of
Fig~\ref{fig:ener}.  Furthermore, these distributions are shifted with
respect to one another.

\begin{figure}
\hspace{0.15cm}
\includegraphics[width=82mm,clip]{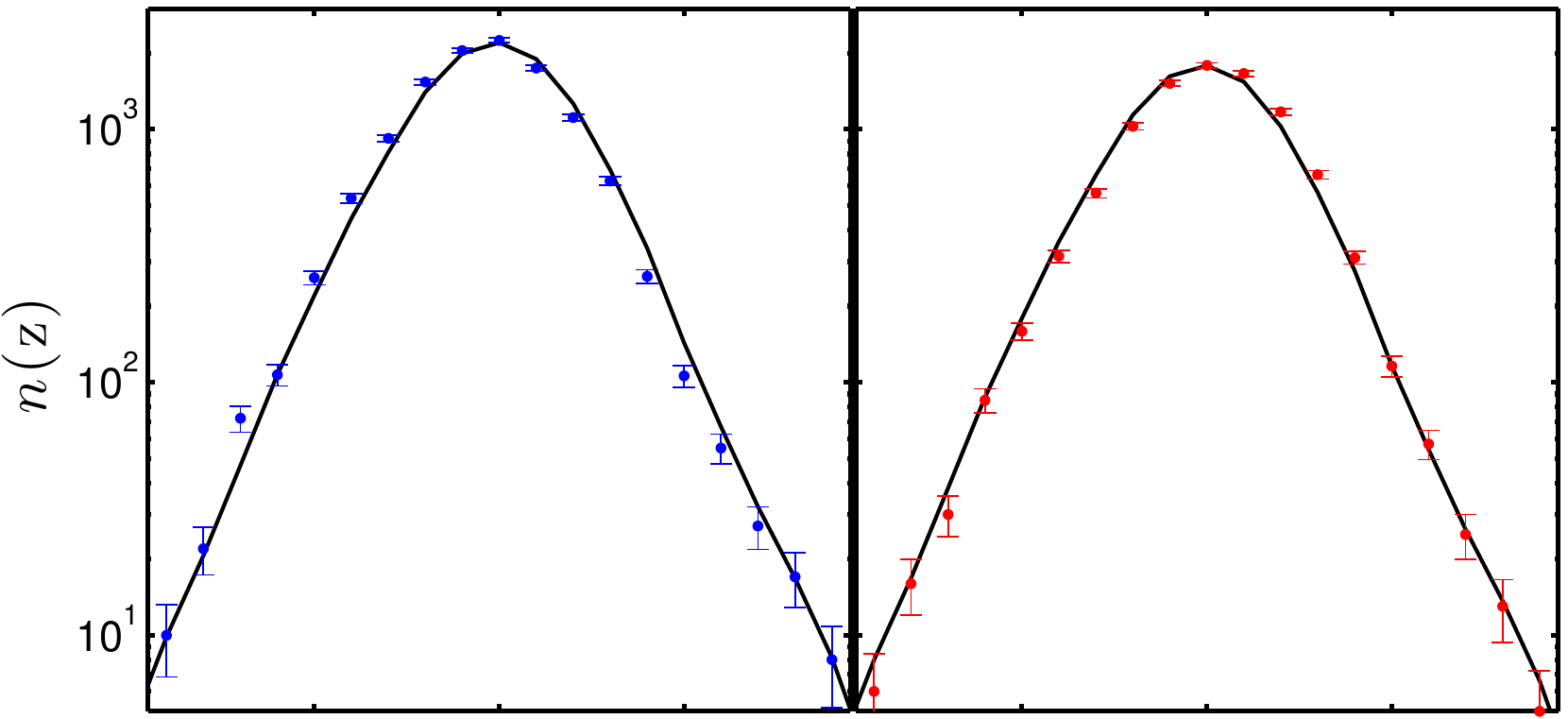}\\
\includegraphics[width=84.3mm,clip]{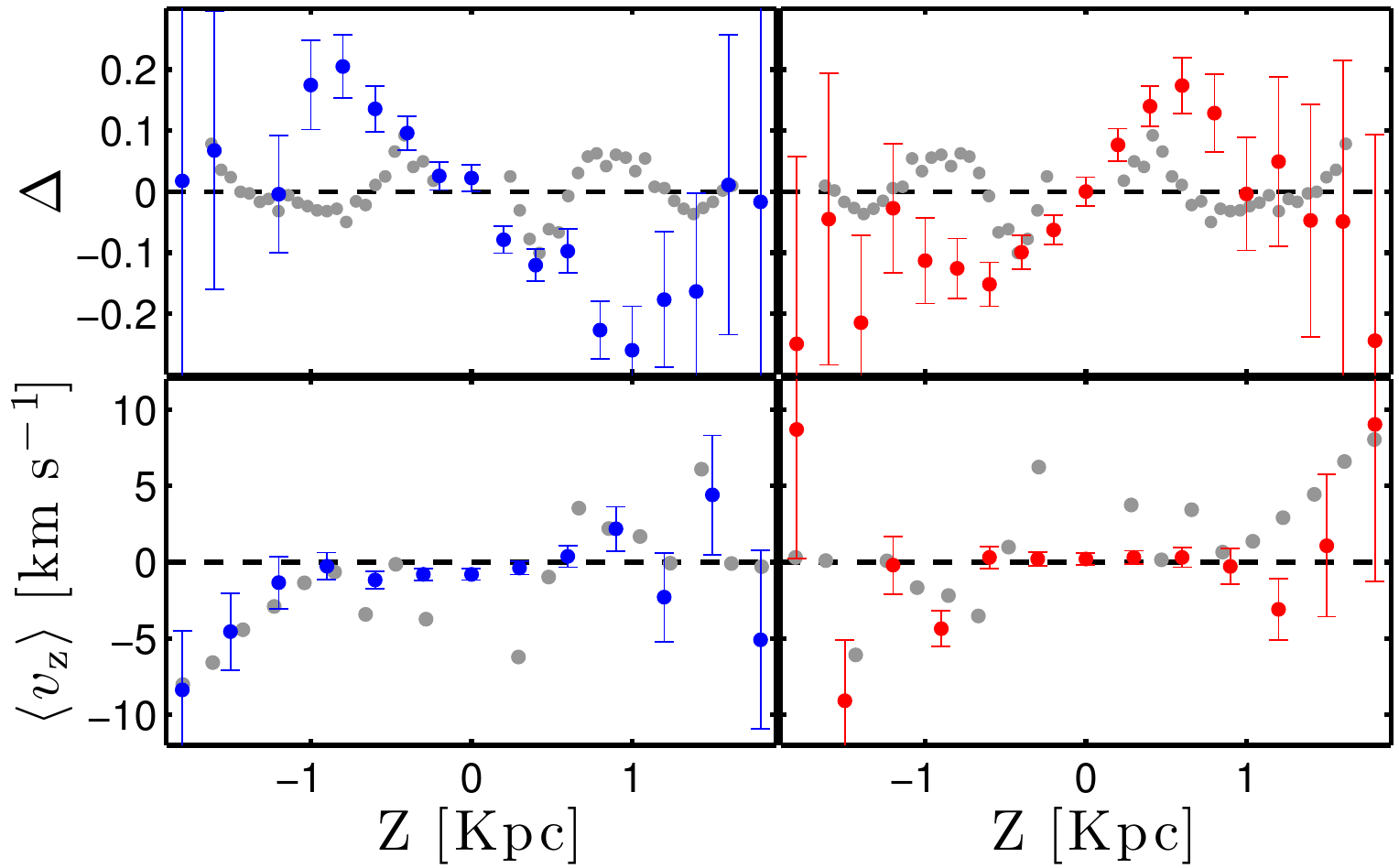}
\caption{{\it First row}: The blue  and red dots show the distribution
  of particles as a function  of height, $n({\rm Z})$, with respect to
  the  midplane  of the  disc,  obtained  from  the volumes  shown  in
  Fig~\ref{fig:dens}.   The black  solid  lines show  a  model of  the
  underlying smooth  distribution $n_{\rm av}({\rm  Z})$.  {\it Second
    row}:  Distribution of  the  residuals, $\Delta  =  (n({\rm Z})  -
  n_{\rm av}({\rm  Z}))/n_{\rm av}({\rm  Z})$, for both  volumes. {\it
    Third row}:  Mean vertical velocity  as a function Z.   Error bars
  indicate Poisson  noise. The grey dots  show observational data
    from  W12.   On  the  right-hand   panels  we  have   shifted  the
    obervational data  with respect to  the axes Z  = 0 and $v_{\rm  z} =
    0$ to match  the phases  of the  waves.}
\label{fig:widrow}
\end{figure}

\begin{figure}
\includegraphics[width=84.3mm,clip]{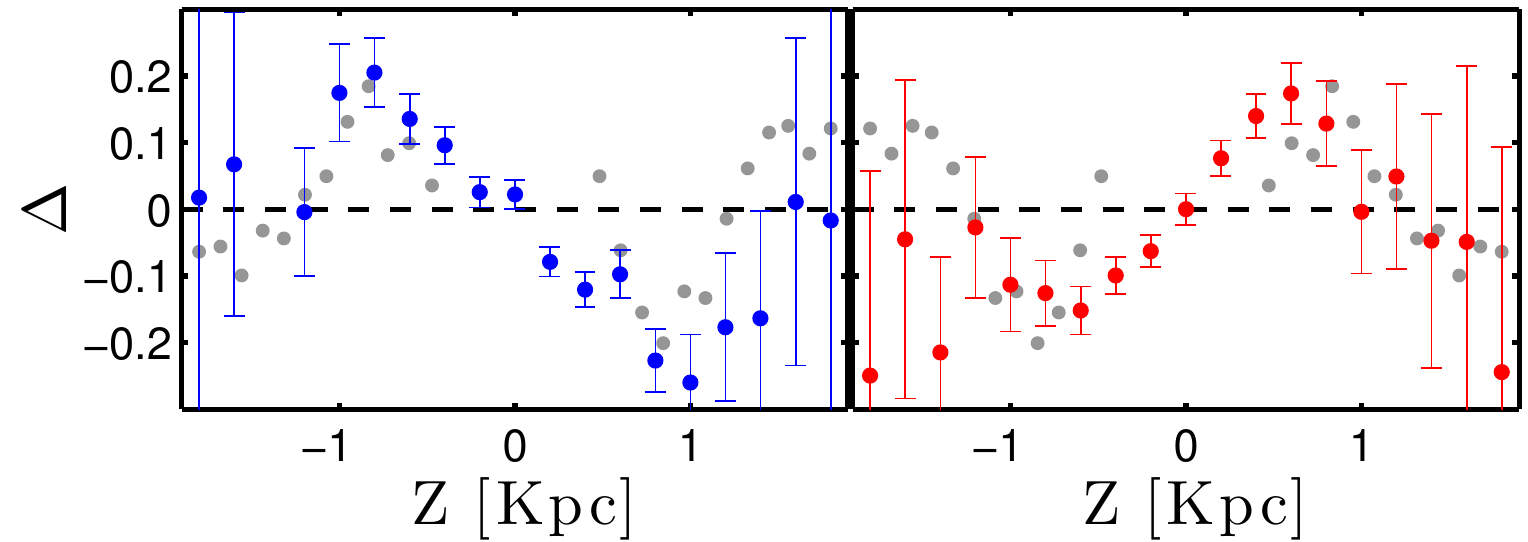}
\caption{As  in   Fig.~\ref{fig:widrowb},  after  rescaling  the
    amplitude and  extent of the observed  $\Delta$ by a  factor of 2.
    On the right-hand panel we have shifted the obervational data with
    respect to the axis  Z = 0 to match the phases  of the waves. Note
    the very similar wave-like  behavior seen in the simulated and
    observed data sets.}
\label{fig:widrowb}
\end{figure}

To explore whether the asymmetries  and shifts observed in the $E_{\rm
  z}$ distributions are  indications of \emph{vertical} density waves,
we compute for both volumes  the distribution of stellar particles as
a  function of height  with respect  to the  midplane of  the galactic
disc,  $n({\rm  Z})  _{|_{R,\theta}}$,  or  simply  $n({\rm  Z})$.   A
north-south asymmetry in this distribution could be an indication of
vertical modes (see  W12). To compute $n({\rm Z})$,  we have carefully
aligned  the disc with  the X-Y  plane.  This  is done  by iteratively
computing,  and aligning  with  the Z-direction,  the total  angular
momentum of the  disc particles located within 4  kpc radius cylinders
of  decreasing height.  In  order to  identify signatures  of vertical
density waves, it  is desirable to compare $n({\rm  Z})$ obtained from
the  perturbed  disc  with  its corresponding smooth underlying
distribution.   For this  purpose, W12  fitted a  smooth two-component
model to their stellar sample's number density. In this work we follow
a  different  approach: we  obtain  a  \emph{smooth} distribution  of
stellar particles,  as a function of height,  by azimuthally averaging
$n({\rm     Z})$    i.e.,     
\begin{equation}
\nonumber
n_{\rm    av}({\rm     Z})_{|_{R}}    = (2\pi)^{-1}\int_{0}^{2\pi}{n(Z,\theta)_{|_{R}}~d\theta}.
\end{equation}  
Our  assumption is  that local  asymmetries of  this  distribution are
erased  after averaging over  all azimuthal  angles.  In  addition, we
expect $n_{\rm av}({\rm Z})_{|_{R}}$  to be a better representation of
the true smooth height  distribution of particles than smooth analytic
fits.   The top  panels of  Fig.~\ref{fig:widrow} show,  with coloured
dots, the  $n({\rm Z})$ distributions  obtained from the  ``blue'' and
``red'' volumes  in Fig.~\ref{fig:dens}, whereas the  black solid line
corresponds to $n_{\rm av}({\rm Z})$.  Note that $n_{\rm av}({\rm Z})$
was re-normalised to the total number of particle in each volume.  Due
to finite particle numbers, we are able to reliably track $n({\rm Z})$
only  up to $|{\rm  Z}| \approx  1.4$ kpc.   These panels  indicates a
shift of the  local with respect to the  smooth distribution, although
this occurs in different directions for the two cases.  Note, however,
that for  large $|Z|$  (i.e. $\gtrsim  0.7 $ kpc),  the shift  in both
distributions  becomes progressively  smaller, exhibiting  a wave-like
pattern.  As in W12, we plot the residual, 
\begin{equation}
\nonumber
\Delta = \dfrac{n({\rm Z})
  -  n_{\rm av}({\rm  Z})}{n_{\rm av}({\rm  Z})},
\end{equation}  
to highlight  these asymmetries.  This is  shown in the  second row of
Fig.~\ref{fig:widrow} with blue and red dots.  In both cases, $\Delta$
is an  odd function of Z.   The grey dots  show data from Figure  1 of
W12, derived  from a  sample of main-sequence  stars in  SDSS-DR8, the
Eighth Data  Release of the  Sloan Digital Sky Survey  \citep{DR8}. To
match  the phases  of  the waves,  on  the right-hand  panels we  have
shifted the obervational data with respect  to the axis ${\rm Z} = 0$.
At least within  $|{\rm Z}| < 1.4$ kpc,  the residuals associated with
our volumes present  a similar wave-like behavior to  that observed in
the SN.  However, the amplitude  and extent of the perturbations found
in this simulation are larger by  a factor of $\sim 2$. This indicates
that either the Heavy Sgr model  is too massive or that we are looking
at the  perturbation at an earlier  stage. In addition, the  lack of a
gaseous  disc component  in  these  simulations is  likely  to play  a
significant role.  A fraction of  the energy imparted by the satellite
should be  absorbed by the  gas, thus weakening the  perturbation.  In
Fig.~\ref{fig:widrowb},   we  compare   the  observed   and  simulated
$\Delta$, after rescaling the  observational data by the aforemetioned
factor.  Note the very similar  behavior of $\Delta$ observed in these
two data  sets.  In the  bottom row of Fig.~\ref{fig:widrow},  we show
$\langle v_{\rm z} \rangle$ as a  function of Z.  We note that results
associated with this distribution should  be taken with caution, due to
their relatively  low statistical significance.  As  before, grey dots
indicate data  extracted from the  top panel of  Figure 4 in  W12.  At
large  Z ($\approx  1$ kpc),  the  ``blue'' (``red'')  volume shows  a
global  trend  towards  negative  (positive)  values  as  Z  decreases
(increases).  As explained by W12,  this suggests a coherent motion of
stellar particles away from the disc midplane.

It is  important to realise  that the phase-space distribution  of the
“blue” SN-like  volume qualitatively reproduces not  only the observed
vertical structure  (W12), as we just demonstrated,  but also features
seen in the  plane of the local Galactic disk: As  shown by G12, local
samples of old disc stars  present a total energy distribution that is
in good  agreement with  the distribution shown  in the left  panel of
Fig. ~\ref{fig:ener} (blue lines).

\begin{figure*}
\centering
\includegraphics[width=84.6mm,clip]{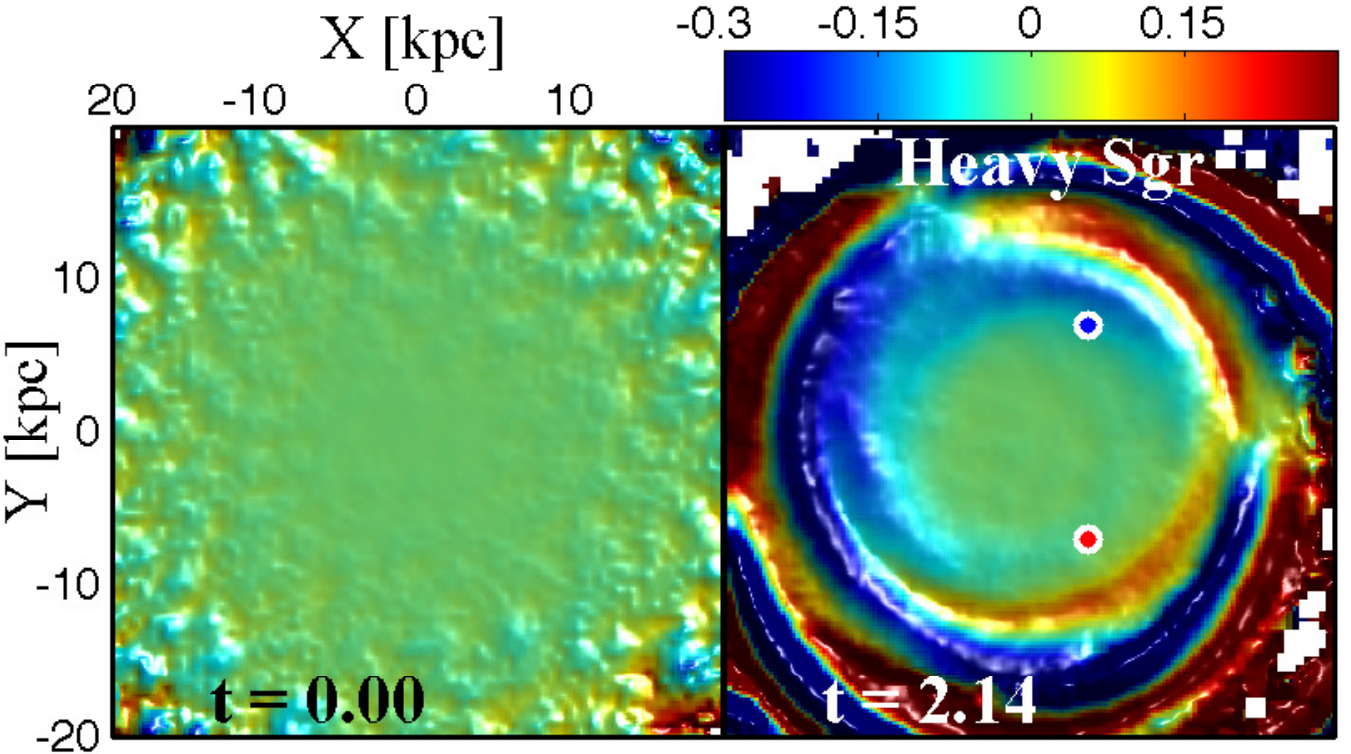}
\hspace{-0.3cm}
\includegraphics[width=77.7mm,clip]{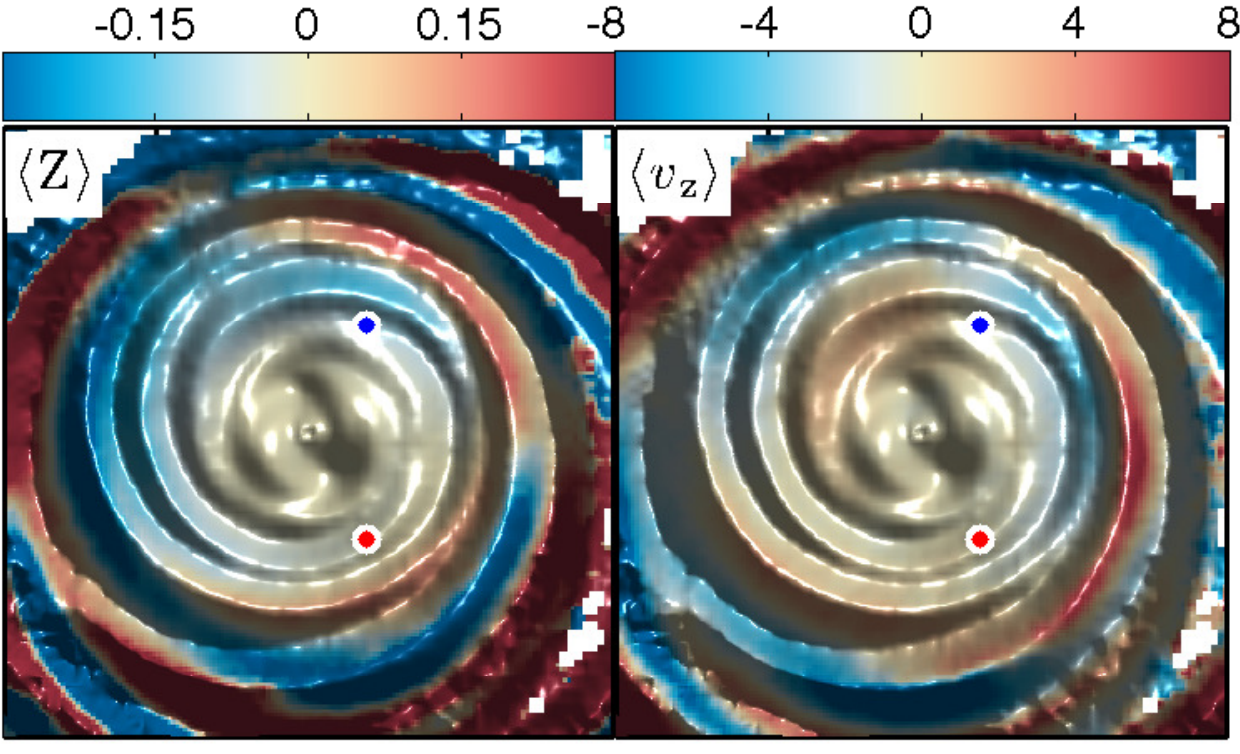}
\\
\hspace{0.49cm}
\includegraphics[width=78.28mm,clip]{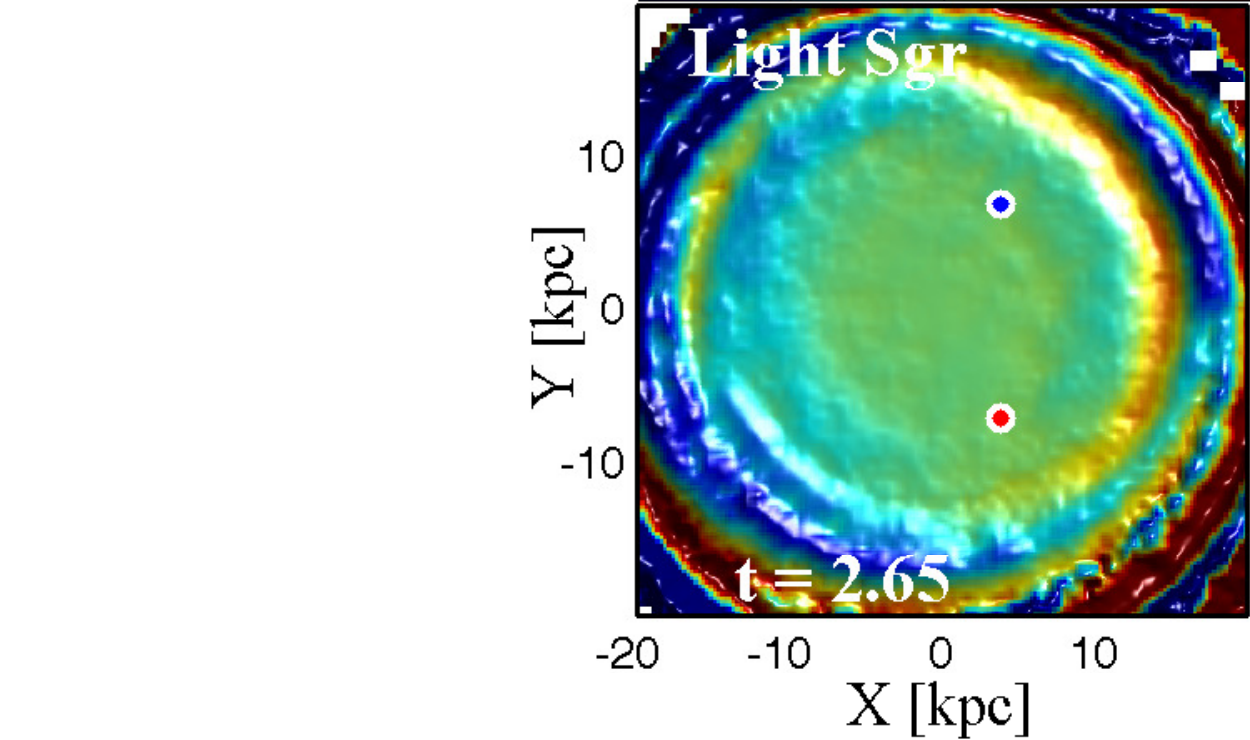}
\hspace{-0.2cm}
\includegraphics[width=78.5mm,clip]{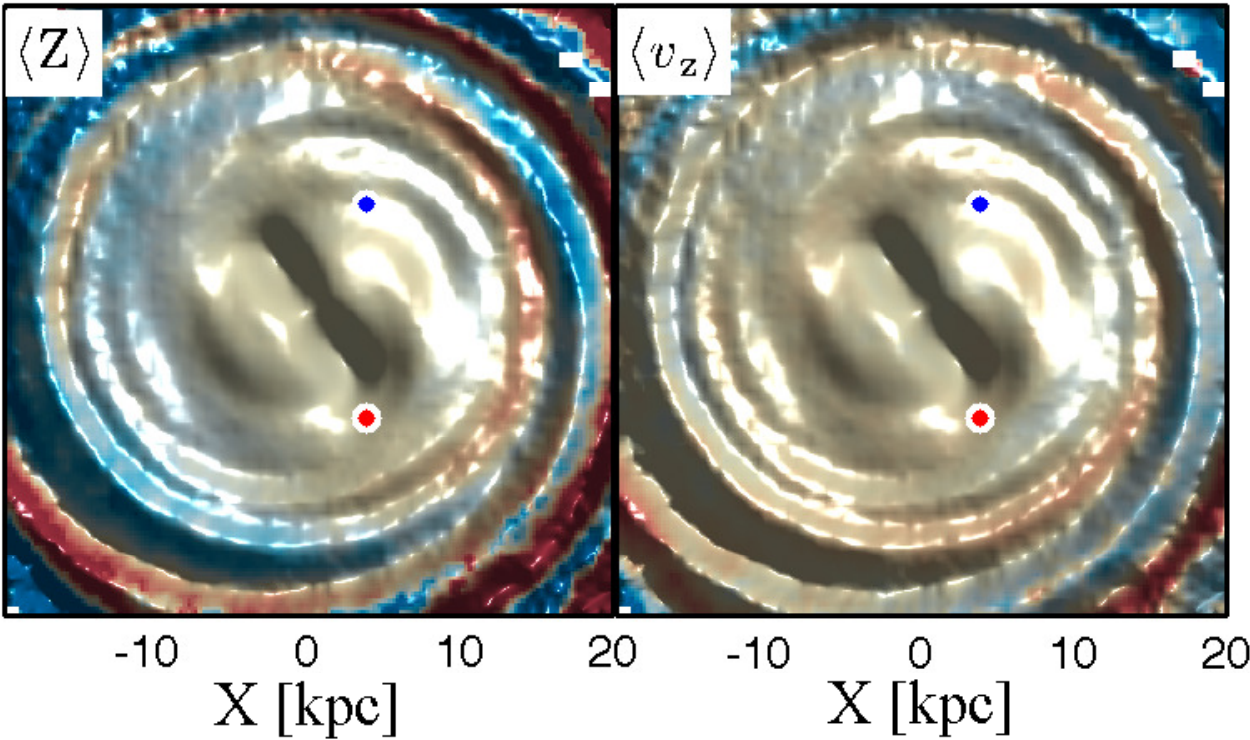}

\caption{{\it First column}: Map of the simulated galactic disc's mean
  height,  $\langle {\rm  Z}  \rangle$,  at $t=0$  Gyr,  used in  both
  simulations out  to $R=20$  kpc.  {\it Second  column}: Maps  of the
  simulated galactic  disc's $\langle {\rm Z}  \rangle$ at present-day
  configuration, obtained from the  Heavy (top) and Light (bottom) Sgr
  simulations,  respectively.    The  different  colours   and  relief
  indicate different values of $\langle {\rm Z} \rangle$ in kpc.  {\it
    Third  column}:  Different colours  indicate  different values  of
  $\langle  {\rm Z}  \rangle$ in  kpc, whereas  the relief  traces the
  overdensity maps shown  if Fig.~\ref{fig:dens}. {\it Fourth column}:
  As before, but for the mean vertical velocity (in km s$^{-1}$).  The
  coloured dots in the top  panel indicate the location of the volumes
  analysed in Sec.~\ref{sec:local}.}
\label{fig:maps}
\end{figure*}

\section{Azimuthal and radial dependance}
\label{sec:global}

\begin{figure}
\centering
\includegraphics[width=80mm,clip]{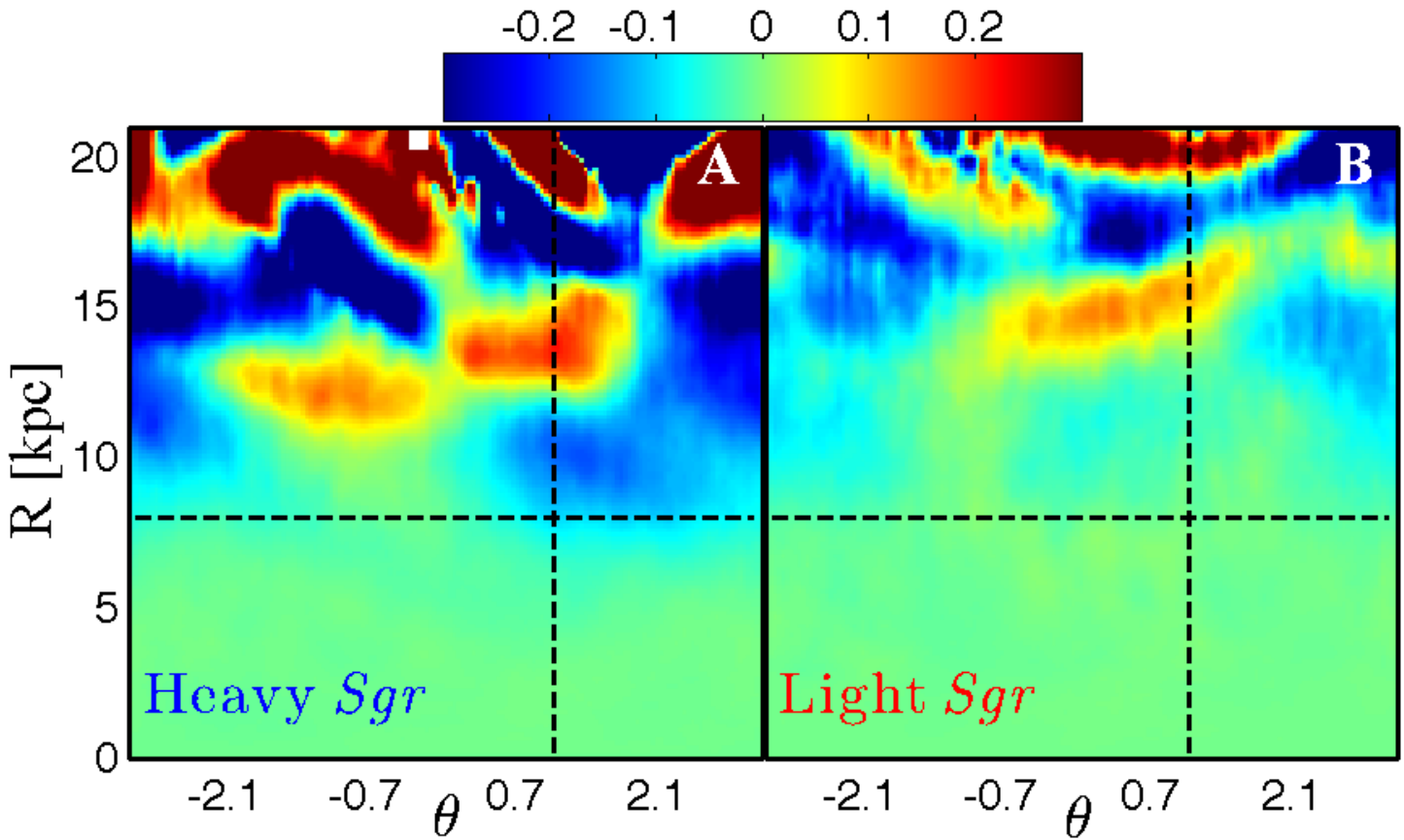}\\
\hspace{0.4cm}
\includegraphics[width=60mm,clip]{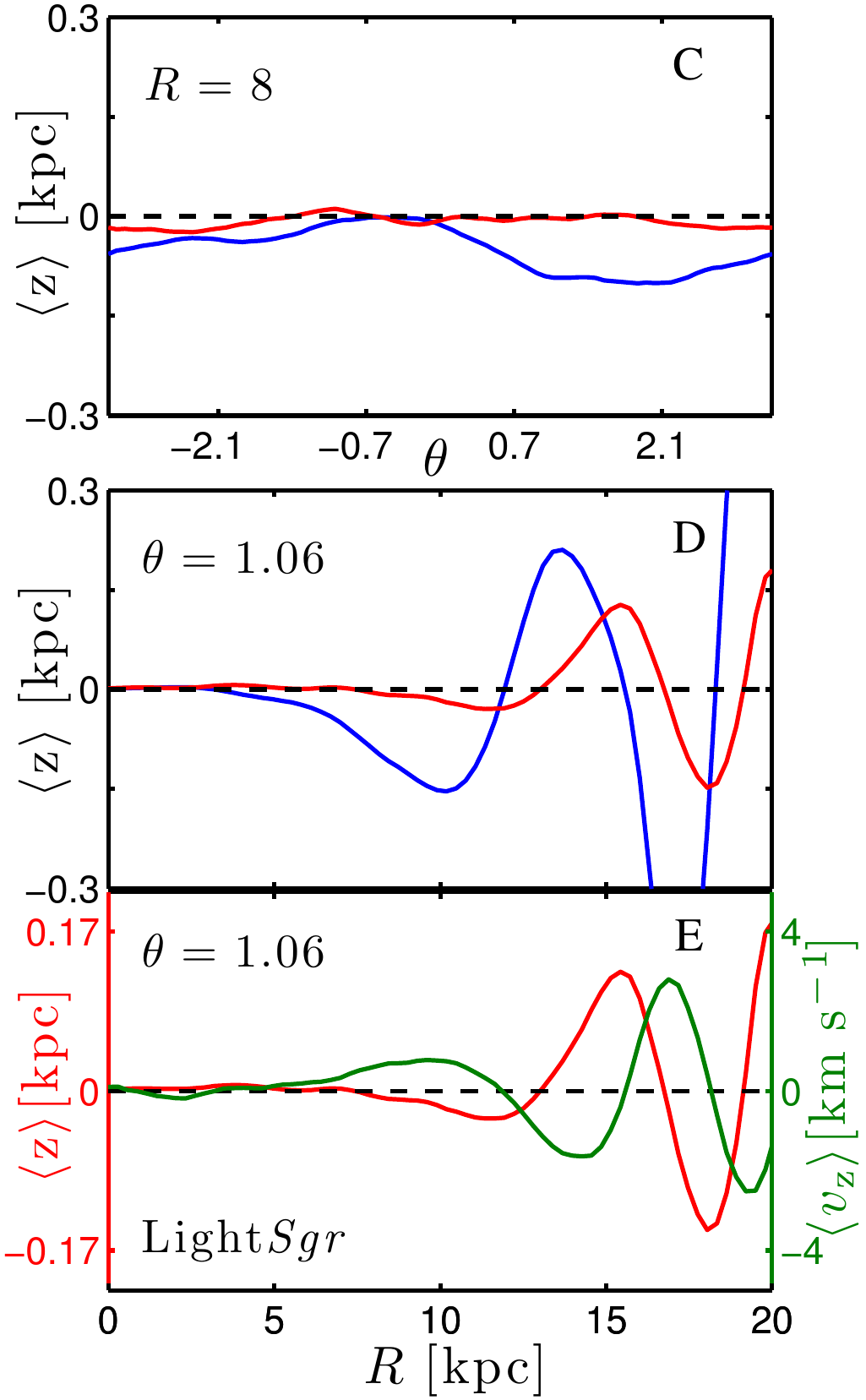}
\caption{{\it Panels A and B}:  Maps of the simulated galactic disc's,
  $\langle  {\rm Z}  \rangle$, at  present-day  configuration obtained
  from the  Heavy (left)  and Light (right)  {\it Sgr}  simulations in
  polar coordinates (in kpc).  The black dashed  lines cross at the location of
  the blue volume  shown in Fig~\ref{fig:dens}.  {\it Panels  C and D}
  show the variation of the mean  height of the disc as we move across
  the black dashed lines shown in  the top panels.  Blue and red lines
  are  associated  with the  Heavy  and  Light  {\it Sgr}  simulation,
  respectively.  {\it  Panel E}: Comparison  of the mean  height (red)
  and  vertical  velocity (green),  as  a  function of  galactocentric
  radius obtained from the Light {\it Sgr} simulation.}
\label{fig:polar}
\end{figure}

We   have  shown   that   the  phase-space   distributions  of   Solar
Neighbourhood-like  volumes   extracted  from  our   Heavy  {\it  Sgr}
simulation exhibit wave-like  perturbations in the vertical direction.
However,  it is not  clear from  the analysis  shown thus  far whether
these are localised  perturbations or are signatures of  a global mode
perturbing the entire disc.   From Fig.~\ref{fig:widrow}, we know that
the $n({\rm Z})$ distributions of the two analysed volumes are shifted
with  respect to  one  another, as  well  as with  respect to  $n_{\rm
  av}({\rm Z})$.   If the observed  shifts are signatures  of vertical
waves, we would expect to find  correlations in the mean height of the
disc $\langle {\rm Z} \rangle$  as a function of galactocentric radius
and azimuthal angle.  We explore this in Fig.~\ref{fig:maps}, where we
show maps  of $\langle {\rm Z}  \rangle$ for the Heavy  and Light {\it
  Sgr} simulation  disc out  to $R \approx  20$ kpc.  To  obtain these
maps,  we grid  the disc  with a  regular Cartesian  mesh of  bin size
$=0.5$ kpc aligned with the X-Y  plane.  On each grid node we centre a
1 kpc radius cylinder and compute $\langle {\rm Z} \rangle$ by fitting
a Gaussian  distribution to $n({\rm Z})$.   In the top  left panel, we
show  the  $\langle  {\rm   Z}  \rangle$-map  reflecting  the  initial
conditions of both discs at  $t=0$ Gyr.  As expected from an initially
unperturbed  disc,  the maps  are  consistent  with  $\langle {\rm  Z}
\rangle =  0$ at all radii,  except for small-scale  departures due to
finite particle  resolution.  The situation  dramatically changes when
we  explore  the present-day  configurations.   The  second column  of
panels in Fig.~\ref{fig:maps} shows  the $\langle {\rm Z} \rangle$-map
of the  Heavy (top) and  Light (bottom) {\it Sgr}  simulation's discs,
obtained  after $t  \approx  2.1$ and  $2.6$  Gyr, respectively.   The
colour coding indicates different  values of $\langle {\rm Z} \rangle$
and white regions  represent volumes devoid of particles.   We can now
clearly identify wave-like,  spiral perturbations traveling across the
disc.  It  is possible to appreciate  how the mean height  of the disc
gradually  increases (decreases) as  we follow  one of  these patterns
azimuthally.  Departures of $\langle  {\rm Z} \rangle$ with respect to
the midplane can be as large as 0.3 kpc, especially for the Heavy {\it
  Sgr} simulation (top panels).  The local volumes analysed previously
are    indicated    with   coloured    dots.     As   expected    from
Fig.~\ref{fig:widrow}, the  blue (red) volume  is located in  a region
lying slightly  below (above) the  midplane.  In this  simulation, the
inner 7 kpc of the disc  has not been strongly vertically perturbed by
the satellite galaxy.  Although the Light {\it Sgr} simulation (bottom
panels)  also exhibits  wave-like perturbations,  they  generally have
smaller amplitudes.  Note that  strong perturbations are observed only
at $R  \gtrsim 10$ kpc. In  both cases, such boundary  is located near
corrotation, where density waves  are reflected backwards to the outer
disc.   This can  be more  clearly appreciated  in panels  A and  B of
Fig.~\ref{fig:polar}, where we show the $\langle {\rm Z} \rangle$-maps
in polar  coordinates.  The  black dashed  lines, at $R  = 8$  kpc and
$\theta  =  1.06$  rad  (measured  with  respect  to  the  positive  X
semi-axis), cross at the location of the blue volume.  For comparison,
this  location is  indicated in  both  discs.  Notice  that the  inner
regions of the Light {\it Sgr}  disc presents a value of $\langle {\rm
  Z}  \rangle \approx  0$  for a  larger  radial extent.   Panel C  in
Fig.~\ref{fig:polar} shows $\langle {\rm Z} \rangle$, as a function of
$\theta$,  at   $R=8$  kpc  for  both   discs.   Significant  vertical
perturbations at  this galactocentric radius  can be observed  only in
the  Heavy {\it  Sgr} simulation.   A dependance  of $\langle  {\rm Z}
\rangle$  with azimuthal  angle $\theta$  is present.   Panel  D shows
$\langle {\rm Z} \rangle$, as a function of $R$, at $\theta=1.06$ rad.
In both discs, a well-defined  wave-like pattern can be observed, with
an amplitude increasing as a function of $R$.  Note that $\langle {\rm
  Z}  \rangle$  exhibits  a  stronger dependance  with  galactocentric
radius than  with azimuthal angle;  a result that could  be contrasted
against currently  available stellar samples.  In the  third column of
Fig.~\ref{fig:maps},  we   explore  the  relationship   between  these
vertical patterns  and the radial modes  shown in Fig.~\ref{fig:dens}.
Here, the different colours indicate different values of $\langle {\rm
  Z} \rangle$, whereas the relief traces overdense regions.  Overdense
features can be found both above and below the midplane.  Furthermore,
we can appreciate how the mean height of a given spiral arm changes as
a function of the azimuthal angle.

If  the observed patterns  are indeed  signatures of  vertical density
waves, a  correlation between $\langle  {\rm Z} \rangle$ and  the mean
vertical velocity $\langle v_{\rm z} \rangle$ is expected.  We explore
this in panel E of  Fig.~\ref{fig:polar}, where we plot, for the Light
{\it  Sgr} disc,  $\langle {\rm  Z}  \rangle$ and  $\langle v_{\rm  z}
\rangle$ as a function of $R$, at a fixed azimuthal angle.  A wave-like
pattern  is   also  observed   in  $\langle  v_{\rm   z}\rangle$.   At
galactocentric radii where $\langle {\rm Z} \rangle$ takes an extrema,
$\langle  v_{\rm  z}\rangle  \approx   0$.   On  the  other  hand,  at
galactocentric radii where $\langle v_{\rm z} \rangle$ takes a maximum
or  a minimum value,  $\langle {\rm  Z} \rangle  \approx 0$.   This is
exactly  what is  expected from  oscillatory behaviour.   In  the last
column  of Fig.~\ref{fig:maps}  we show,  with different  colours, the
local $\langle  v_{\rm z}\rangle$ for both galactic  disc.  The relief
in these panels traces the corresponding overdensity maps.  Note again
the wave-like  structure of  these patterns.  In  the Heavy  {\it Sgr}
simulation (top panel), $\langle v_{\rm z}\rangle$ can depart from $0$
km s$^{-1}$ by more than 8 km s$^{-1}$.  Similarly to what is observed
for $\langle  {\rm Z}  \rangle$, the $\langle  v_{\rm z}\rangle$  of a
given spiral arm varies as a function of the azimuthal angle, changing
from positive  to negative departures with respect  to $\langle v_{\rm
  z}\rangle = 0$ km s$^{-1}$.

\section{The Magellanic clouds as other possible culprit}

  We have focused our attention on perturbations induced by the Sgr
  dwarf galaxy.   However, the Large Magellanic Cloud  (LMC) have been
  considered in the  past by several authors as  a plausible perturber
  behind  the  warp  observed  in  the MW  disk's  ${\rm  H_I}$  layer
  \citep{h1warp}.  Given  its traditional mass  ($\sim 2\%$ of  the MW
  mass) and location ($\sim 50$ kpc) estimates, the LMC tidal field is
  not sufficiently strong to induce the observed vertical perturbation
  \citep[e.g.][hereafter B07]{huntertoomre,besla07}.  Nonetheless, the
  addition of the force from the  dark matter halo wake excited by the
  Clouds \citep{weing98}  could be enough to account  for the observed
  warp  \citep{weing06}.   The   previous  results,  based  on  linear
  perturbation   theory,  were   tested  with   fully  self-consistent
  simulations by \citet{tsu}.  Assuming  a decaying orbit for the LMC,
  this analysis showed that the observed warp could be reproduced over
  a 6 Gyr timescale, involving approximately four full orbital periods
  \citep[see  also][for   a  different  opinion]{GR}.    This  orbital
  configuration  is at  odds with  the recent  findings by  B07, which
  suggested that the Magellanic Clouds  are approaching the MW for the
  very  first  time on  a  parabolic  orbit  traveling at  $\sim  400$
  km/s. However, \citet{vespe} showed that the perturbation excited by
  a  low-velocity ($\sim 200$  km/s) flyby  encounter with  a slightly
  more massive satellite ($ > 5\%$  of the MW mass) can be efficiently
  transmitted to the  inner regions of the dark  matter halo, where it
  can affect the  structure of an embedded stellar  disk.  More recent
  estimates  of  the  total  LMC  mass, based  on  abundance  matching
  techniques,  suggested values  as large  as  $10\%$ of  the MW  mass
  \citep{bkb}.  Moreover,  the orbits of  the MCs are  currently being
  revised (Kallivayalil et al., in preparation).  New dynamical models,
  including both updated orbital  parameters and total mass estimates,
  are  required  to  assess   whether  such  a  mechanism  would  have
  sufficient time to operate. We defer this analysis to future work.

\section{Discussion}
\label{sec:disc}

In this  work, we  have explored  a scenario in  which Sgr  is the
perturber behind  the North-South  asymmetry recently observed  in the
number density and mean vertical velocity of Solar Neighbourhood stars
(W12). For  this purpose, we have  searched for both  local and global
signatures of vertical density  waves in two simulations modelling the
response of  the MW  to the infall  of Sgr.  Distributions  of stellar
particles as a function  of height in Solar Neighbourhood-like volumes
extracted from  our more  massive Sgr's progenitor  simulation present
clear indications of  a perturbation in the vertical  direction of the
disc.  This asymmetry  becomes more evident when comparing  to a model
of the  underlying smooth vertical distribution  of particles.  Within
the  $|{\rm Z}|$-range  allowed  by our  finite  mass resolution,  the
phase-space distribution of  certain SN-like volumes can qualitatively
reproduce the North-South asymmetry  observed by W12.  Remarkably, the
same  phase-space  distributions   can  simultaneously  reproduce  the
signatures of radial density waves observed  by G12.

By creating  maps of the  mean height of  the disc within $R  \leq 20$
kpc, we have  shown that the vertical perturbations  observed in local
volumes are signatures of a global mode perturbing the entire disc. As
in the case  of the radial density modes (G12),  the amplitude and the
extent to  which vertical modes  can radially penetrate into  the disc
depends on  the mass of  the perturbing satellite.   Interestingly, we
have shown that  the mean height of the disc is  expected to vary much
more    rapidly    in   the    radial    than    in   the    azimuthal
direction. Furthermore,  the mean height of  overdense spiral features
vary azimuthally, moving from below to above the midplane of the disc.
Signatures of vertical modes should  also be observable in maps of the
Galactic disc's  mean vertical velocity since,  not surprisingly, they
present a clear oscillatory behavior.

In  contrast to  radial  modes that  can  be excited  by  a number  of
different external  and internal mechanisms, vertical  modes {\it must
  be excited} by some agent external to the disc. Altough we have
  shown that perturbations  induced by Sgr could be  enough to account
  for  several features  observed  in the  Solar  Neighborhood, it  is
  likely  that MCs  are  playing  a  role  in shapping  the
  vertical structure of MW disc.  We plan to characterize the coupling
  of these  two perturbations in a follow-up  study. Contrasting the
results presented in this  work against currently available samples of
Galactic  disc  stars could  help  to  understand  the origin  of  the
observed vertical and in-plane perturbations.

\section*{Acknowledgments}
We would like  to thank the anonymous referee  for the useful comments
and suggestions which helped to improve this paper.  The authors wish to
thank  Gurtina Besla  for insightful  discussions.  FAG  was supported
through the  NSF Office  of Cyberinfrastructure by  grant PHY-0941373,
and  by  the Michigan  State  University  Institute for  Cyber-Enabled
Research.  BWO was  supported  in  part by  the  Department of  Energy
through the  Los Alamos  National Laboratory Institute  for Geophysics
and Planetary  Physics.  TCB  acknowledges partial support  from grant
PHY  08-22648: Physics  Frontiers Center/Joint  Institute  for Nuclear
Astrophysics (JINA), awarded by the U.S. National Science Foundation.

\bibliographystyle{mn2e}
\bibliography{vertical_waves}

\label{lastpage}
\end{document}